\documentstyle[12pt,epsf]{article}
\def\lsim{\mbox{\,\raisebox{.3ex}{
           $<$}$\!\!\!\!\!$\raisebox{-.9ex}{$\sim$}\,}}

\def\0{\over } \def\1{\vec } \def\2{{1\over2}} \def\4{{1\over4}}

\let\a=\alpha  \let\g=\gamma \let\d=\delta
 \let\z=\zeta  
  \let\l=\lambda \let\m=\mu
 \let\s=\sigma
 \let\o=\omega

  \let\D=\Delta
\def\wu#1{\sqrt{{#1} \,}^{ \hbox to0.2pt{\hss$
     \vrule height 2pt width 0.6pt depth 0pt $} \;\! } }
\def\P{{\mit \Pi}} \def\mn{{\mu\nu}}

\def\({\left(} \def\){\right)} \def\<{\langle } \def\>{\rangle }
\def\[{\left[} \def\]{\right]} \def\lb{\left\{} \def\rb{\right\}}
\let\lk=\[     \let\rk=\]

\def\be{\begin{equation}}    \def\ee{\end{equation}}
\def\bea{\begin{eqnarray}}   \def\eea{\end{eqnarray}}
\def\nn{\nonumber\\}

\catcode`\@=11
\long\def\@makefntext#1{
\protect\noindent \hbox to 3.2pt {\hskip-.9pt
$^{{\ninerm\@thefnmark}}$\hfil}#1\hfill}           

\def\@makefnmark{\hbox to 0pt{$^{\@thefnmark}$\hss}}  

\def\ps@myheadings{\let\@mkboth\@gobbletwo
\def\@oddhead{\hbox{}
\rightmark\hfil\ninerm\thepage}
\def\@oddfoot{}\def\@evenhead{\ninerm\thepage\hfil
\leftmark\hbox{}}\def\@evenfoot{}
\def\sectionmark##1{}\def\subsectionmark##1{}}

\renewcommand{\thefootnote}{\fnsymbol{footnote}}

\newcounter{sectionc}\newcounter{subsectionc}
                     \newcounter{subsubsectionc}
\renewcommand{\section}[1] {\vspace*{0.6cm}
                     \addtocounter{sectionc}{1}
\setcounter{subsectionc}{0}\setcounter{subsubsectionc}{0}\noindent
        {\normalsize\bf\thesectionc. #1}\par\vspace*{0.4cm}}
\renewcommand{\subsection}[1] {\vspace*{0.6cm}
             \addtocounter{subsectionc}{1}
        \setcounter{subsubsectionc}{0}\noindent
        {\normalsize\it\thesectionc.\thesubsectionc. #1}\par
        \vspace*{0.4cm}}
\renewcommand{\subsubsection}[1]
{\vspace*{0.6cm}\addtocounter{subsubsectionc}{1}
        \noindent
{\normalsize\rm\thesectionc.\thesubsectionc.\thesubsubsectionc.
        #1}\par\vspace*{0.4cm}}

\newcounter{appendixc}
\newcounter{subappendixc}[appendixc]
\newcounter{subsubappendixc}[subappendixc]

\renewcommand{\appendix}[1] {\vspace*{0.6cm}
        \refstepcounter{appendixc}
        \setcounter{figure}{0}
        \setcounter{table}{0}
        \setcounter{equation}{0}
        \renewcommand{\thefigure}{\Alph{appendixc}.\arabic{figure}}
        \renewcommand{\thetable}{\Alph{appendixc}.\arabic{table}}
        \renewcommand{\theappendixc}{\Alph{appendixc}}
        \renewcommand{\theequation}{\Alph{appendixc}.\arabic{equation}}
        \noindent{\bf Appendix \theappendixc #1}\par\vspace*{0.4cm}}

\def\abstracts#1{{

\centering{\begin{minipage}{12.2truecm}\footnotesize
      \baselineskip=12pt\noindent
        \centerline{\footnotesize ABSTRACT}\vspace*{0.3cm}
        \parindent=0pt #1
        \end{minipage}}\par}}

\renewenvironment{thebibliography}[1]
        {\begin{list}{\arabic{enumi}.}
        {\usecounter{enumi}\setlength{\parsep}{0pt}
\setlength{\leftmargin 1.25cm}{\rightmargin 0pt}
         \setlength{\itemsep}{0pt} \settowidth
        {\labelwidth}{#1.}\sloppy}}{\end{list}}

\newcounter{itemlistc}     \newcounter{romanlistc}
\newcounter{alphlistc}     \newcounter{arabiclistc}

\newcommand{\fcaption}[1]{
    \refstepcounter{figure}
    \setbox\@tempboxa = \hbox{\footnotesize Fig.~\thefigure. #1}
        \ifdim \wd\@tempboxa > 6in
           {\begin{center}
        \parbox{6in}{\footnotesize
        \baselineskip=12pt Fig.~\thefigure. #1} \end{center}}
        \else
             {\begin{center}
             {\footnotesize Fig.~\thefigure. #1}
              \end{center}}
        \fi}

\newcommand{\tcaption}[1]{
    \refstepcounter{table}
    \setbox\@tempboxa = \hbox{\footnotesize Table~\thetable. #1}
    \ifdim \wd\@tempboxa > 6in
       {\begin{center}
    \parbox{6in}{\footnotesize\baselineskip=12pt
    Table~\thetable. #1}
        \end{center}}
        \else
             {\begin{center}
             {\footnotesize Table~\thetable. #1}
              \end{center}}
        \fi}

\def\@citex[#1]#2{\if@filesw\immediate\write\@auxout
        {\string\citation{#2}}\fi
\def\@citea{}\@cite{\@for\@citeb:=#2\do
        {\@citea\def\@citea{,}\@ifundefined
        {b@\@citeb}{{\bf ?}\@warning
        {Citation `\@citeb' on page \thepage \space undefined}}
        {\csname b@\@citeb\endcsname}}}{#1}}

\newif\if@cghi
\def\cite{\@cghitrue\@ifnextchar [{\@tempswatrue
        \@citex}{\@tempswafalse\@citex[]}}
\def\citelow{\@cghifalse\@ifnextchar [{\@tempswatrue
        \@citex}{\@tempswafalse\@citex[]}}
\def\@cite#1#2{{$\null^{#1}$\if@tempswa\typeout
        {IJCGA warning: optional citation argument
        ignored: `#2'} \fi}}

 \font\ninerm=cmr9

\begin{document}
\begin{flushright}
{\tt
DESY 95-091 \hfill ISSN 0418-9833\\
ITP-UH-17/95 \hfill hep-ph/9505307\\
May 1995 \hfill~\\}
\end{flushright}
\baselineskip=22pt
\centerline{\normalsize\bf HOT SCALAR ELECTRODYNAMICS}
\baselineskip=16pt
\centerline{\normalsize\bf AS A TOY MODEL FOR HOT QCD\footnote{%
Talk given by A.K.R. at
a one-day meeting dedicated to the memory of
Tanguy ALTHERR, held on November 4, 1994 at CERN, Geneva.
To appear in a Gedenkschrift
published by World Scientific.}}

\vspace*{0.6cm}
\centerline{\footnotesize ULRIKE KRAEMMER\,,\, ANTON K. REBHAN}
\baselineskip=13pt
\centerline{\footnotesize\it DESY, Gruppe Theorie, Notkestr. 85,
D-22603 Hamburg, Germany}
\vspace*{0.3cm}
\centerline{\footnotesize and}
\vspace*{0.3cm}
\centerline{\footnotesize HERMANN SCHULZ}
\baselineskip=13pt
\centerline{\footnotesize\it Institut f. Theoret. Physik,
            Universit\"at Hannover,}
\baselineskip=12pt
\centerline{\footnotesize\it Appelstr. 2, D-30167 Hannover,
            Germany}

\vspace*{0.9cm}
\abstracts{
Hot scalar electrodynamics is adopted as a toy model for a hot
gluon plasma to display some aspects of the compulsory
resummation of hard thermal loops when next-to-leading order
quantities at soft momentum scales are to be calculated.}

\normalsize\baselineskip=15pt
\setcounter{footnote}{0}
\renewcommand{\thefootnote}{\alph{footnote}}
\section{Introduction}

One of Tanguy Altherr's major activities was the development
and application of an improved perturbation theory for quantum
field theory at high temperature\cite{TA} and/or high
density\cite{TU}. Although his real interest with respect to
the former was in hot QCD, which will be hopefully probed more
or less directly in heavy-ion collisions in the near future ---
a future, sadly, without Tanguy --- , he naturally was
also interested in studying toy models which can give guidance
in analysing and overcoming the theoretical problems involved.

A particularly simple toy model is provided by self-interacting
scalar fields\cite{TAphi,Parwani}, in which some of the issues
associated with the resummation of hard thermal
loops\cite{BP,FT} can already be elucidated. Here we shall
rather take massless scalar fields interacting through
electrodynamics to furnish a toy model which is closer to the
self-interacting massless gauge bosons of QCD. Notice that
spinor QED is a much less interesting toy model since the
infrared behaviour of fermions is much milder on account of
Pauli suppression. Our toy model Lagrangian thus is
\be
 {\cal L} = (D_\mu\phi)^* D^\mu\phi-\textstyle{1\over4}
 F_{\mu\nu}F^{\mu\nu}
\ee
with $D_\mu=\partial_\mu+ieA_\mu$.

Focussing on the perturbative corrections to the thermal photon
self-energy, we shall discuss to what extent the properties of
hot gluons and the required resummation methods can be
understood already in the Abelian toy model.

\section{Hard thermal loops}

On dimensional grounds, in the limit of high temperature $T$
much larger than any momentum or mass scale, the photon
self-energy, as well as the one for the scalar particles, will
receive one-loop contributions proportional to $e^2T^2$. Hence,
this will give rise to important modifications of the spectrum
on momentum scales $Q_0,q\lsim eT$.

Let us begin with the simpler scalar self-energy. In the
imaginary-time formalism, the one-loop self-energy is given by
the usual Feynman integrals with the difference that the
integral over frequencies is replaced by a sum over imaginary
Matsubara frequencies $\o_n=2\pi n T$,
\be  \label{Sigma}
  \Sigma = e^2 T\sum_n\int{d^3k\0(2\pi)^3} \({3\0K^2}-
  Q^2  {3-\a \0 K^2 (K-Q)^2}
    + { 2(\alpha -1) K\cdot Q \0 K^4 (K-Q)^2 } \) \;\; ,
\ee
where $\a$ is the gauge fixing parameter of general covariant
gauges. At first, this is defined only for discrete imaginary
$Q_0$, but can be continued analyticly to real continuous
frequencies $Q_0$.

In the limit $T\gg Q_0,q$, only the first term in (\ref{Sigma})
turns out to contribute, which happens to be gauge-parameter
independent. Evaluating the thermal sum through contour
integrals\cite{Kapu} one finds
\bea  \label{mu2}
  e^2 T\sum_n\int{d^3k\0(2\pi)^3}{3\0K^2} &=& 3e^2
  T\sum_n\int{d^3k\0(2\pi)^3} {-1\0(2\pi nT)^2+k^2} \nn
  &=& -3e^2\int{d^3k\0(2\pi)^3}{n(k)+\2\0k} \nn
  &=& -\4e^2T^2 \;\equiv\; -\mu^2
\eea
upon dropping an UV-divergent temperature-independent
contribution (or using dimensional regularisation), which
would be cancelled by the usual zero-temperature mass
counterterm.

The result (\ref{mu2}) shows that the original massless scalar
particles acquire a certain {\em thermal mass}. It should be
borne in mind, however, that this is {\em not} quite the same
as a zero-temperature rest mass! For example, calculating the
finite-temperature corrections to the energy-momentum tensor,
one finds that it remains traceless (up to the usual
zero-temperature trace anomaly).

Turning to the less simple photon self-energy, one first of all
faces a more complicated tensorial structure because the
existence of a preferred frame, the plasma rest-frame, `breaks'
Lorentz symmetry. Actually this means simply that there is an
additional vector $U_\mu$ ($=\d_\mu^{\;0}$ in the plasma
rest-frame), to build covariant quantities with. In particular
there exists now a vector which is transverse to a given
momentum $Q_\mu$,
\be
   V_\mu=Q^2 \, U_\mu-(U\cdot Q) Q_\mu \;\; ,
\ee
and this allows one to construct a novel transverse symmetric
Lorentz tensor
\be
   P_\ell=V\otimes V/V^2 \;\; .
\ee
Together with the usual transverse projector one can build one
which is also spatially transverse and orthogonal to $P_\ell$,
\be
   P_t = g-{Q\otimes Q\0Q^2}-P_\ell \;\; .
\ee

The photon self-energy $\P^\mn(Q)$, which is transverse with
respect to $Q_\mu$, therefore involves two (rather than the
usual one) structure functions,
\be
   \P^\mn=\P_t P_t^\mn+\P_\ell P_\ell^\mn \;\; .
\ee

In the nonabelian case this turns out to be more complicated.
Indeed, already the one-loop contribution beyond the leading
$T^2$ part is nontransverse in gauges other than
Feynman\cite{EHKT,KKM}. The Abelian case is also special in
that the gauge-boson self-energy is manifestly gauge parameter
independent. At one-loop order it reads
\be\label{Pimn}
   \P^\mn(Q) = e^2 T\sum_n\int{d^3k\0(2\pi)^3} \[
   {(2K-Q)_\mu(2K-Q)_\nu\0 K^2 (K-Q)^2}
     -{ 2 g_\mn\0K^2} \] \;\; .
\ee
In the limit $T\gg Q_0,q$ one obtains
\bea \label{plhtl}
   \P_\ell & = & -{Q^2 \0 q^2} \P_{00} = 3m^2 \( 1
      -{Q_0^2 \0 q^2} \) \(1 - {Q_0 \0 2q} \ln{Q_0+q\0Q_0-q}\) \\
 \label{pthtl}
    \P_t & = & {1\02}\(3m^2-\P_\ell\) \;\; ,
\eea
with $m=eT/3$. As concerns this leading temperature correction,
the nonabelian gluon self-energy differs only in that one has
to replace $e^2\to g^2(N+N_f/2)$ for SU($N$) with $N_f$
fermions\cite{KalKlW}.

Unlike the scalar self-energy, there is no simple thermal mass
term but $\P_{t,\ell}$ are nontrivial functions of $Q_0/q$.
Correspondingly, the poles in the propagator which describe the
normal modes of the plasma are not given by a simple mass
hyperboloid, but are given by the transcendental equations
$Q^2=\P_{\ell,t}(Q)$. The resulting dispersion laws
$Q_0=\o_{\ell,t}(q)$ for the photonic excitations are given in
Fig.~1. One finds that the mass $m=eT/3$ introduced above equals
the minimum frequency for which propagating modes ($q^2>0$,
{\it i.e.} $q$ real) exist. Whereas the longitudinal branch
approaches the light-cone as $q$ is increased, the `effective
mass' of the transverse branch grows with a limiting value
$eT/\wu 6$. Also given in Fig.~1 are the poles of the photon
propagator for negative $q^2$. These do not correspond to
normal modes, but describe the response of the system to
localised perturbations with given frequency component $\o<m$.
Such perturbations are screened exponentially with screening
length $|q(\o)|$. In the static limit, only the longitudinal
branch gives a finite screening length $1/m_D, m_D=eT/\wu 3 $,
which is the familiar Debye screening of electrostatic fields,
whereas the transverse branch does not, corresponding to
unscreened magnetostatic fields.

In the Abelian case, the absence of a magnetostatic screening
mass can be proved to hold exactly\cite{Fradkin}, whereas in
the nonabelian case, it is known to cause the breakdown of
perturbation theory. The common expectation, first put forward
by Linde\cite{Linde}, is that a magnetic mass is generated
nonperturbatively at order $g^2T$.

 \begin{figure}   \baselineskip=13pt                       
     \centerline{ \epsfxsize=4in                           
     \epsfbox[70 240 540 560]{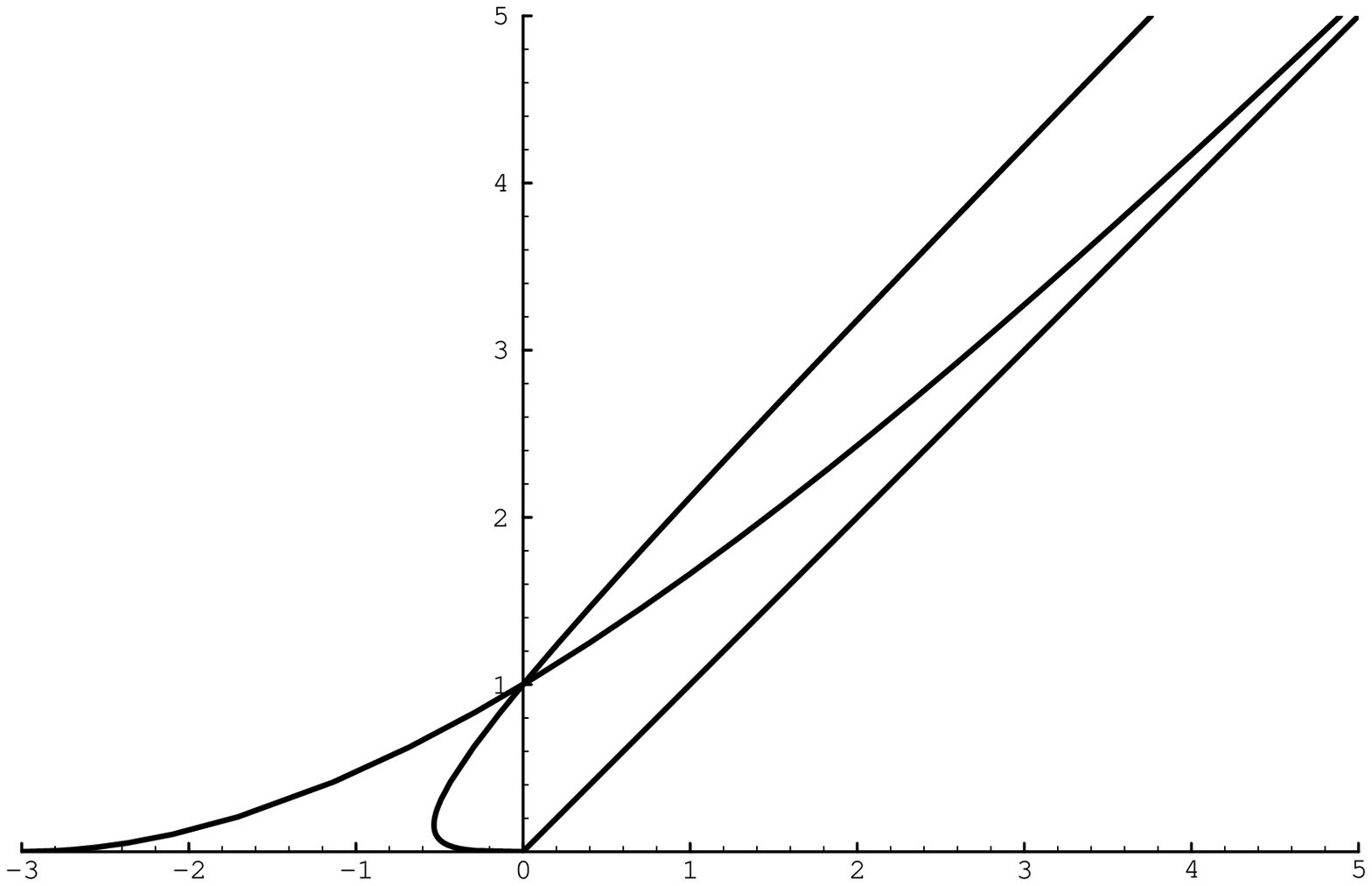} }                      
\unitlength1cm \begin{picture}(0,0)
\put(13.3,0.5){\footnotesize $q^2/m^2$}
\put(4.8,6.8){\footnotesize $\o^2/m^2$}
\put(9,5.3){\footnotesize $(t)$}
\put(10,5.15){\footnotesize $(\ell)$}
\put(10,4){\footnotesize $\nwarrow$}
\put(10.5,3.8){\footnotesize \it light-cone}
\put(6.45,0){$\longmapsto$ {\footnotesize \it plasma waves}}
\put(2.65,0){{\footnotesize \it dynamical screening} $\longleftarrow$}
\end{picture}\\
\fcaption{The dispersion laws of transverse ($t$) and longitudinal
($\ell$) photonic (gluonic) quasiparticles}
\end{figure}                                               

\section{The need for resummation}

Hard thermal loops, the simplest of which is the thermal mass
acquired by scalar particles, Eq.~(\ref{mu2}),
typically\footnote{~In renormalizable theories in four
dimensions. The hard thermal loops for gravitons\cite{RFT},
for example, are proportional to $T^4$.}\
{}~have integrands which
would diverge quadratically in the ultraviolet, if there was no
exponential cutoff at the scale $T$ through the distribution
function $n$. Hence the term `hard thermal' --- the dominant
scale for the loop momentum is $k\sim T$.

The subleading terms in the high-temperature expansion,
however, begin to probe smaller momenta, and finally run into
infrared problems when bosons are around, for in the infrared,
the bosonic distribution function diverges, $n(k)\sim T/k$
for $k\ll T$.

These infrared problems reflect a breakdown of the usual loop
expansion. When loop integrals are dominated by the momentum
scale $eT$ (or $gT$ in QCD), the thermal masses, which are
one-loop objects, become as important as the tree-level kinetic
term. Hence, pertuurbation theory breaks down unless these
particular one-loop quantities are treated on equal footing with
the bare Lagrangian.

In order to rescue perturbation theory one has to resum the
hard thermal loops. This can be achieved by adding them to the
bare Lagrangian. To avoid overcounting, they have to be
subtracted again as `thermal counterterms', as which they are
treated as one-loop objects. This makes sure that the original
theory has not changed. This rearrangement of perturbation
theory can also be understood as resulting from integrating out
in a first step all the hard modes with momentum $k\ge\l$, with
$\l$ such that $gT\ll\l\ll T$, which produces an effective
Lagrangian containing the hard thermal loops. In a second step,
the soft modes with $k\le\l$ are covered, but using the
effective Lagrangian.

In general, there are not only thermal masses which are
comparable to bare quantities at soft momentum scales, but
also hard thermal vertices. In QED and QCD there are in fact
infinitely many of those, which have been first classified
by Frenkel, Taylor, Braaten, and Pisarski\cite{FT,BP}. The
simplicity of scalar field theory is that there are
none\cite{Parwani}, which also holds true for scalar
electrodynamics\cite{KRS}.

The hard thermal self-energies of scalar electrodynamics can
be summarized by the following Lagrangian,
\be
   {\cal L}^{\rm HTL} = - \mu^2 \phi^\ast \phi
    + {3 \0 4} m^2 \int{d\Omega\04\pi} Y^\rho F_{\rho \mu} \,
    { 1 \0 (Y \partial )^2 } \, F^{\mu \lambda} Y_\lambda \;\; ,
\ee
where $Y^\mu$ is a light-like vector with $Y^0=1$. For the
scalars, there is just a simple mass term, whereas the photons
have a nonlocal effective Lagrangian, corresponding to the more
complicated results in Eqs.(\ref{plhtl},\ref{pthtl}). In QED as
well as in QCD, similar nonlocalities appear, but because of
gauge invariance they involve $[Y\cdot D(A)]^{-1}$, which
generates hard thermal vertices with an arbitrary number of
external gauge boson lines\cite{eff,shortcut,BPeff}.

\section{$\P_\mn$ beyond leading order}

The leading-order result for the polarization tensor is given
by the hard thermal loops (\ref{plhtl},\ref{pthtl}),
$\P_\mn\sim e^2T^2 \sim m^2$. The next term in the
high-temperature expansion of the bare one-loop result is
$\P_\mn(Q)\sim e^2QT\sim em^2$ for $Q_0,q\sim m$, and, were it
not for the breakdown of perturbation theory discussed above,
two-loop contributions should start with $e^4T^2\sim e^2m^2$
and be negligible. However, in the bare theory the $\ge2$-loop
contributions are increasingly infrared-divergent, and also
contribute to the relative order $e$ upon resummation. In the
following we shall look at the effects of this resummation in
some simple cases.

\subsection{Static limit of $\P_\ell$ -- electrostatics}

Before resummation, the one-loop result for $\P_\ell(Q_0=0,q)$
reads
\be \label{plunr}
  \P_\ell(0,q) = \P_{00}(0,q)={e^2T^2\03}
 + {e^2\024\pi^2} q^2 \ln{\s\0T}+ \ldots
 = 3m^2 \(1+ O(e^2) \) \;\; ,
\ee
where $\s$ is the renormalization scale, and one would not
expect corrections to the Debye screening mass $m_D=\wu 3 m$
which are larger than $O(e^2)$. Resumming the thermal mass of
the scalars, however, changes the one-loop result to\cite{KaKli}
\bea\label{md}
  \P_{00}(0,q)&=&{e^2T^2\03} - {e^2T\0\pi^2} \int_0^\infty \! dk \,
  \lb { \m^2 \0 k^2+\m^2 } + 1 - {k\0q} \ln
  \left| {2k+q\02k-q} \right| \rb + O(e^2 q^2 \ln(T)) \nn
  &=&
  {e^2T^2\03} - {e^2T\mu\02\pi} + O(e^2 q^2 \ln(T)) \nn
  &=&3m^2\(1-{3\04\pi}e+O(e^2)\).
\eea

This result can in fact be derived in a simplified resummation
scheme, which has been put forward by Arnold and
Espinosa\cite{AE}: in the imaginary time formalism it is clear
that only the static modes need to be resummed; the nonstatic
modes have frequencies $\sim T$, so self-energy corrections
$\sim e^2T^2$ can be treated perturbatively. Keeping only the
static modes one sees at once that the result (\ref{md}) has
to be momentum independent at relative order $e$, since only
the last term in Eq.~(\ref{Pimn}), which corresponds to the
seagull diagram, contributes.

In QCD the self-interacting bosons analogously give rise to
relative-order-$g$ corrections, but there $\d\P_{00}(0,q)$
does not happen to be a constant. $\d\P_{00}(0,q)$, and also
$\d\P_{00}(0,q\to0)$, is gauge dependent, even after
resummation\cite{Toimela}, which for some time was taken as
indication that the nonabelian Debye mass cannot be extracted
from the gluon propagator\cite{Nadkarni}. The
resolution\cite{AKR} is that $\d\P_{00}(0,q)$ has to be
evaluated at the location of the pole in the leading-order
gluon propagator, which is at $q=\pm i m_D$. There a gauge
independent result can be extracted, which however turns out
to be logarithmically infrared divergent unless a magnetic
screening mass is assumed.

\subsection{Static limit of $\P_t$ -- magnetostatics}

The unresummed one-loop result for $\P_t(Q_0=0,q)$, which
pertains to the magnetostatic sector, reads
\be \label{piiunr}\P_t(0,q)=-\2
  \P_{ii}(0,q) = {1\016}e^2qT + {e^2\024\pi^2}q^2\ln{\s\0T}
     + \ldots = {3\016}mqe \( 1 +O(e) \) \;\; .
\ee

Since this vanishes for $q\to0$, there appears to be no
generation of a magnetic mass, but if Eq.~(\ref{piiunr})
was correct, this would imply a space-like pole in the
magnetostatic propagator
\be
  \D_t(0,q) = {-1\0q^2-e^2qT/16} \;\; ,
\ee
at $q=e^2T/16$. Static magnetic fields would not decay
monotoneously but rather oscillate in space.

However, resummation changes all that and instead yields
\be \label{piir}
   \P_{ii}(0,q)={e\m\02\pi} \lk 2\mu-{q^2+4\m^2\0q}
   \arctan\(q\02\m\) \rk + O(e^2 \mu^2) \;\; .
\ee
The result (\ref{piiunr}) turns out to be correct only in the
limit $q\gg\mu\sim eT$, whereas for $q\to0$ the true behaviour
is $\propto q^2$. This can in fact proved to hold in all
orders\cite{BIP}. Thus there is indeed no magnetic screening
mass and also no space-like poles in the magnetostatic
propagator.

In QCD the unresummed result is similar to (\ref{piiunr}), but
there the resummation of hard thermal loops only modifies the
position of the space-like pole, without completely removing
it. The latter is sometimes called the Landau ghost of thermal
QCD and it is assumed that the nonperturbative generation of
a magnetic mass is what will eventually remove it.

\subsection{Long-wave-length limit -- plasma frequency}

In the limit $q\to0$ with $Q_0\not=0$ the poles in the
transverse and in the longitudinal component of the photon
propagator coincide and determine the plasma frequency
$\o_{\rm pl.}\equiv m=eT/3$, above which there are propagating
normal modes of the plasma.

In the unimproved one-loop result the subleading terms are
given by
\be \label{plwlunr}
  \P_{t,\ell}(Q_0,0) = {e^2T^2\09}-{e^2T\012\pi}iQ_0
  - {e^2\024\pi^2}Q_0^2 \ln{\s \0 T} + O(e^2 Q_0^2 T^0)
  = m^2-{e\04\pi}iQ_0m + O(e^2 m^2) \;\; .
\ee
The next-to-leading order term now is purely imaginary,
apparently implying a non\-zero damping constant
\be \label{gunr}
 \g = -{1 \0 2m} {\rm Im}\;\P_{t,\ell}(Q_0=m,0)
    = {e^2T\024\pi} = {e\08\pi}m \;\; .
\ee

But after resummation, this changes completely:
\be \label{plwlr2}
 \P_t(Q_0,0)=\P_\ell(Q_0,0)={e^2T^2\09}+{e^2T\02\pi}\lb
 -\m-{4\03Q_0^2}\( [\m^2-(Q_0/2)^2]^{3\02}-\m^3 \) \rb \;\; .
\ee
Now this is real at $Q_0=m$. The unresummed result is seen to
be correct only for $Q_0\gg m$ and its imaginary part is due to
the fact that with $Q_0\gg m$ one is above the threshold for
pair production of scalar quasi-particles, whereas $m<2\mu$.

The analogous calculation in QCD is equally misleading prior
to resummation of hard thermal loops. There the bare one-loop
damping constant comes out even gauge dependent and negative,
which has caused a lot of confusion for quite some
time\cite{samm}, and was in fact the driving force for the
development of the Braaten-Pisarski resummation scheme.

After resummation, the QCD result does give a nonzero (and
positive and gauge-independent) damping constant\cite{BPdamp},
which is in fact due to higher-order Landau damping rather
than quasi-particle pair production.

In scalar electrodynamics, the result (\ref{plwlr2}) implies
only a correction to the plasma frequency,
\be \label{5.37}
  m^2+\d m^2={e^2T^2\09}\(1-{8\wu2-9\02\pi}e\)\approx
  {e^2T^2\09}(1-0.37 e) \;\; .
\ee
The corresponding calculation in QCD has been performed in
Ref.~\cite{HS}, yielding
$$ (\d m^2/m^2)_{\rm QCD} \approx -0.18  \wu {g^2N} \;\; . $$
Let us see how far this latter result can be understood by the
above result on scalar electrodynamics. From the leading order
terms it is clear that $e^2$ corresponds to $g^2N$, so we might
try to apply (\ref{plwlr2}) by inserting the gluonic plasmon
mass in place of the thermal mass of the scalars. This would
give $(\d m^2/m^2)\approx -0.028 \wu{g^2N}$, which is over a
factor of 6 short of the actual result. Hence, the correction
to the QCD plasma frequency is much larger than what might be
expected from just the appearance of thermal masses in the loop
integrals.

Another point worth mentioning is that the result (\ref{plwlr2})
is not obtained correctly when only the static modes are
resummed. Keeping only the static modes and trying an analytic
continuation of the result afterwards gives
\bea \label{plwlstr}
   \d\P_{t,\ell}(Q_0,0)\Big|_{\rm static \, contr.}
   &=&-{1\03}e^2T\int{d^3k\0(2\pi)^3}\lb
   {4k^2\0 (k^2+\m^2)(k^2+\m^2-Q_0^2)}-{6\0k^2+\m^2} \rb \nn
   &=& -{e^2T\06\pi}\lb {2\0Q_0}\lk {\m\0Q_0}
   - \wu{ {\m^2\0Q_0^2}-1 } \,\rk (Q_0^2-\m^2)+\m \rb \;\; .
\eea
which obviously disagrees with (\ref{plwlr2}).

The pitfall is that the separation of the zero modes relies on
the imaginary time formalism where either $Q_0=0$ or
$Q_0\propto T$, {\it i.e.} hard. Thus, a continuation to soft
$Q_0\not=0$ is precluded.

\section{Resummation close to the light-cone}

As a final limiting case we shall consider $Q^2\to0$, which
is relevant for corrections to the longitudinal branch of the
dispersion laws for large $q$ (see Fig.~1).

Close to the light-cone the leading order (hard thermal loop)
result diverges logarithmically,
\be \label{pihtllc}
  \P_{00}^{\rm HTL}\to-{3m^2\02}\ln{q^2\0Q^2}\quad
  \mbox{for $Q^2\ll q^2$} \;\; ,
\ee
which causes the longitudinal branch to approach the light
cone exponentially without ever piercing it. The corresponding
residue in the longitudinal photon propagator decays
exponentially, too, so that this mode is effectively removed
from the spectrum for large momenta.

Calculating the next-to-leading order contribution to $\P_\mn$
through a resummation of the thermal mass of the scalars (which
can in fact be done for general $Q_0$, $q$ in terms of
elementary functions\cite{KRS} --- albeit with some tedium),
one finds that
\be
   \d\P_{00}\to+e{\mu^2q\0\wu {Q^2} }\quad
   \mbox{for $Q^2\ll q^2$} \;\; .
\ee

This result implies that the light-cone is approached even
quicker\cite{KRS}, but there comes a point where the calculation
can no longer be trusted. With $Q^2\to0$, $\d\P_{00}$ diverges
stronger than $\P_{00}^{\rm HTL}$, so that eventually
$|\d\P_{00}|>|\P_{00}^{\rm HTL}|$ for any arbitrarily small
but nonzero value of $e$ --- perturbation theory breaks down
again.

The origin of this new desaster is in fact already visible in
Eq.~(\ref{pihtllc}). There is a logarithmic singularity at the
light cone which would be lifted by any finite mass for the
hard modes generating the hard thermal loop. In fact, in higher
orders hard lines will also have repeated insertions of hard
thermal loops. Usually, corrections to the hard lines can be
treated perturbatively, but the singular behaviour in the
vicinity of the light cone spoils this. In the case of scalar
electrodynamics, where the only hard thermal loops are
self-energy corrections, it is simple to do a further
resummation of the hard thermal loops to be inserted in hard
lines. This makes a difference only close to the light cone
where $|Q^2|\ll m^2$. With massive internal lines, one can
indeed put $Q^2=0$ and perform a high-temperature expansion
which yields a finite result,
\be \label{pllcres}
  \lim_{Q_0\to q}{\P_{00}^{\rm resum.}} = -{e^2T^2\0 3}
  \lk \ln{2T\0 \mu}+{1\0 2}-\gamma_E+{\z'(2)\0 \z(2)} \rk
  - {e^2T\mu\0 2\pi q^2}+O(e^2 q^2 T^0)
\ee
with $\g_E$ being Euler's constant and $\z$ the Riemann zeta
function.

With Eq.~(\ref{pllcres}), there is now a solution to the
equation $Q^2=\P_\ell=-(Q^2/q^2)\P_{00}$ in the limit
$Q^2\to0$ for a finite value of $q$,
\be \label{qcrit}
  q^2_{\rm crit.}/(eT)^2 = {1\0 3} \lk \ln{4\0 e}
  + {1\0 2}-\gamma_E+{\z'(2)\0\ \z(2)} \rk
  - {e\0 4\pi} = {1\0 3}\ln{2.094\ldots\0 e}-{e\0 4\pi} \;\; .
\ee
Hence, the longitudinal dispersion curve does hit the light
cone. In fact, it continues also somewhat to space-like momentum.

For space-like momentum, there is strong Landau damping from
\be
   \Im m\P_{00}^{\rm HTL}=\theta(-Q^2){3\pi m^2 Q_0\02q} \;\; ,
\ee
but the sharp discontinuity at the light-cone is also an
artefact. After resummation of the hard lines one finds for
$Q^2\ll e^2q^2$
\be
   \Im m\P_{00}^{\rm resum.} =
   {9\pi e^2 m^2 Q_0 \0 16\pi^2 (q-Q_0)}
   \exp\(-e\wu {q\08(q-Q_0)} \)
\ee
so that the imaginary part corresponding to Landau damping
starts from zero with all derivatives vanishing. There is thus
a finite range in $q$ for which weakly damped plasmons with
phase velocity $<1$ exist (the group velocity is $<1$
throughout).

An analogous phenomenon also occurs in hot QCD, which opens
the possibility of \v{C}erenkov interactions\cite{SU}. However,
in the case of QCD, the hard thermal vertices also contribute;
the corresponding calculation still has to be done\cite{FR}.

Another place where the by now well-established resummation
program of Braaten and Pisarski breaks down for similar reasons
is in the case of soft real photon produc\-tion\cite{BPS}.
Again, higher-order corrections to the hard internal lines are
expected to render the result finite\cite{Niegawa}, but a
corrected systematic scheme has still to be developed.

In the toy model of scalar electrodynamics, the essentials of
these at first unexpected problems and their resolution are
already there and give a strong hint how they can be overcome
in the case of hot QCD.

\end{document}